\documentclass{article}%
\usepackage{amsfonts}
\usepackage{amsmath}
\usepackage{amssymb}
\usepackage{graphicx}%
\providecommand{\U}[1]{\protect\rule{.1in}{.1in}}
\newtheorem{theorem}{Theorem}

\newtheorem{remark}[theorem]{Remark}

\newenvironment{proof}[1][Proof]{\noindent\textbf{#1.} }{\ \rule{0.5em}{0.5em}}
\begin{document}

\title{ Maximal entropy distribution functions from generalized R\'{e}nyi entropy}
\author{Gy\"{o}rgy Steinbrecher\\Association EURATOM-MEdC, University of Craiova,\\A. I. Cuza 13, 200585 Craiova, Romania. \\ Email: gyorgy.steinbrecher@gmail.com
\and
Giorgio Sonnino\\Department of Theoretical Physics and Mathematics \\ Universit\'{e} Libre de Bruxelles (ULB)\\ Bvd deTriomphe, Campus Plaine CP 231, 1050 Brussels, Belgium.
\\\&\\
Royal Military School (RMS)\\
Av. de la Renaissance 30, 1000 Brussels, Belgium.\\E-mail: gsonnino@ulb.ac.be
\and
Nicolae Pometescu\\Association EURATOM-MEdC, University of Craiova,\\A. I. Cuza 13, 200585 Craiova, Romania. \\ Email npomet@yahoo.com}
\maketitle

\begin{abstract}
New class of reference distribution functions for numerical approximation of
the solution of the Fokker-Planck equations associated to the charged particle
dynamics in tokamak are studied. The reference distribution functions are
obtained by maximization of the generalized Renyi entropy under
scale-invariant restrictions. Explicit analytic form, with algebraic decay,
that is a generalization of the previous distribution with exponential tails
was derived.

\end{abstract}

\section{Introduction}

The two independently discovered generalizations of the classical
Boltzmann-Shannon Entropy (BSE) \cite{shannon}, by Constantino Tsallis
\cite{Tsallis1}-\cite{TsallisGelMann} respectively by Alfr\'{e}d R\'{e}nyi
\cite{Renyi1}-\cite{Renyi4} found multiple applications \cite{TsallisBook},
\cite{TsallisGelMann}, \cite{Klimontovich}-\cite{RenyiDivergence}. It is a
challenging problem to explain the successes of two similar generalizations of
the BSE. The mathematical "naturalness", of these generalizations was
explained in the framework of Lebesgue $L^{p}$ spaces defined over general
measure space, in the sense that both Tsallis entropy (TE) and the additive
R\'{e}nyi entropy (RE) of order $p$ \ contains the same $L^{p}$ norm for $p>1$
respectively the $L^{p}$ pseudonorm for $0<p<1$ \cite{ReedSimon},
\cite{Rudin}, \cite{LuschgiPages}, \cite{SonninoSteinbrGRE}. A natural
generalization of the R\'{e}nyi entropy (GRE), preserving the its additivity
and geometrical interpretation as a distance in the functions vector space,
where the probability density function is, was given in
\cite{SonninoSteinbrGRE}, \cite{GySGSgrestability}. It was proven
\cite{GySGSgrestability} that by imposing a suitable stabilizing conditions,
the TE, RE and GRE are numerically stable for a large range of the parameters,
they are more stable compared to BSE in the sense that for the stability of
BSE more stabilizing conditions must be imposed \cite{GySGSgrestability}. The
numerical stability of TE, RE and GRE are related to logarithmic convexity
property \cite{GySGSgrestability}. It was proven in that both the TE, RE, GRE,
contains a functional that has good category theoretic properties. The
GRE\ has an interesting application in the study of the complex dynamical
systems \cite{SGYSGstructure}. In a suitable limit the GRE became the RE, that
whose limiting value is the BSE, so the GRE appears as a natural
generalization of the RE, \ for characterization of the singularity or
asymptotic behavior of the multivariate PDF \cite{SonninoSteinbrGRE},
\cite{GySGSgrestability}, or in the characterization of discrete probability
distributions where the set of states is a Cartesian product (the
probabilities has multiple indices) \cite{SGySASGCategory}.

\ The GRE is a Liapunov functional for a large class of dynamical systems
driven by stochastic perturbations \cite{SonninoSteinbrGRE}. Consequently it
is meaningful the study of PDF that realize the maximizes the generalized
entropies. In contrast to the case of BS, RE, TE, in the case of GRE, even in
the simplest case when the total phase space is a Cartesian product of two
smaller spaces, the stationarity condition is expressed by a more complicated
functional equation \cite{SonninoSteinbrGRE}, whose solutions has complex
algebraic decays. The maximal generalized entropy (GMaxEnt) PDF's, whose study
is the object of this article, are interesting because in a series of our
previous works \ \cite{sg1}-\cite{SGSGrRDFderivation}\ we proved that the
reference distribution function (RDF) for charged particle distribution in
tokamak can be obtained by maximization of the BSE, subject to scale
invariant, algebraically the simplest, restrictions. In order to obtain a
better approximation of the PDF of the charged particles in tokamak, in this
work we enlarge the family of RDF obtained previously by MaxEnt principle with
scale invariant restrictions, by considering RDF's obtained from the
maximization of the Generalized R\'{e}nyi entropy.

Reference particle density distribution functions are useful in the numerical
solutions of Fokker-Planck or gyrokinetic equations, that describe charged
particle dynamics in tokamak.

Simplest MaxgEnt distribution functions were studied in
\cite{SonninoSteinbrGRE}. In this article we explore systematically the class
of GRE that appears when the phase space is a Cartesian product of $N=3$ sub
spaces. We use the general formalism of the Lebesgue integration theory that
allows to have an unified formalism for both discrete and continuos
distributions in finite as well as infinite dimensional spaces of stochastic
processes. In the our formalism the R\'{e}nyi divergence appears as a
R\'{e}nyi entropy for a suitable chosen measure \cite{SGySASGCategory}. \ 

\section{T{}he framework, Shannon, R\'{e}nyi and Tsallis entropies}

For starting the discussion about various measures of the information, we need
to specify some exact framework (see \cite{GySGSgrestability} ). We consider a
standard measure space $(\Omega,\mathcal{A},m)$ where $\Omega$ is the phase
space,  $\mathcal{A}$ is the $\sigma$-algebra of the observable events, $m$ is
the measure (that can be discrete, continuos, finite or $\sigma$-finite).  

In the classical definitions with discrete, finite or denumerable probability
space,\ the measure  $m$ is a the counting measure, invariant under
permutation group. In \ many applications when $\Omega$ is a continuum, the
measure $m$ is invariant under physical or geometrical symmetries. In this
article the probability measures
\[
\mathcal{A}\ni A\rightarrow p(A)\in\lbrack0,1]
\]
defined on $(\Omega,\mathcal{A})$  are continuos with respect to measure $m$,
so by Radon-Nicodim theorem we have
\begin{equation}
p(A)=\int\limits_{A}\rho(x)dm(x);~A\subset\Omega
\end{equation}
where $\rho(x)$ is the probability density function. With the previous
notations the Boltzmann-Gibbs-Shannon entropy has the following form
\begin{equation}
S_{cl}[\rho]=-{\int\limits_{\Omega}}\rho(x)\log\left[  \rho(x)\right]  dm(x)
\end{equation}

In the definitions of the  A. R\'{e}nyi \cite{Renyi1} \ respectively by C.
Tsallis \cite{Tsallis1}, \cite{Tsallis2} \ entropies we encounter the same
metric object  in the $L^{p}(\Omega,dm)$ spaces \cite{ReedSimon},
\cite{Rudin}, \cite{LuschgiPages}. For details see ref.
\cite{SonninoSteinbrGRE}. Consequently, with the notations
\begin{align}
\left\Vert \rho\right\Vert _{p} &  =\left[  {\int\limits_{\Omega}}\left[
\rho(x)\right]  ^{p}dm(x)\right]  ^{\frac{1}{p}};~p\geq1\label{LL3}\\
N_{p}[\rho] &  ={\int\limits_{\Omega}}\left[  \rho(x)\right]  ^{p}%
dm(x);~0<p\leq1\label{LL4}%
\end{align}
the entropies of A. R\'{e}nyi \cite{Renyi1} \ $S_{R,q}$\ respectively by C.
Tsallis \cite{Tsallis1}, \cite{Tsallis2} $S_{T,q}$ can be expressed as
follows
\begin{align}
S_{R,q}[\rho] &  =\frac{q}{1-q}\log\left\Vert \rho\right\Vert _{q}%
;q>1\label{LL5}\\
S_{R,q}[\rho] &  =\frac{1}{1-q}\log N_{q}[\rho];0<q<1\label{LL6}\\
S_{T,q}[\rho] &  =\frac{1}{1-q}\left[  1-\left\Vert \rho\right\Vert _{q}%
^{q}\right]  ;q>1\label{LL7}\\
S_{T,q}[\rho] &  =\frac{1}{1-q}\left\{  1-N_{q}[\rho]\right\}
;0<q<1\label{LL8}%
\end{align}
In this formalism the R\'{e}nyi divergence can be expressed as R\'{e}nyi
entropy with suitable chosen measure $dm(\mathbf{x})$ \cite{SGySASGCategory}.

\begin{remark}
\label{markREM_counting_measure}Observe from previous formalism in the case of
discrete probability distribution the original definitions of the R\'{e}nyi or
Tsallis entropies results \cite{Renyi1}, \cite{Tsallis1}.
\begin{align}
S_{R,q}[\rho]  & =\frac{1}{1-q}\log\sum\limits_{k}p_{k}^{q}\\
S_{T,q}[\rho]  & =\frac{1}{1-q}\left[  1-\sum\limits_{k}p_{k}^{q}\right]
\end{align}

\end{remark}

From the previous definitions it is clear that the R\'{e}nyi and Tsallis
entropies are related to the geometric properties Lebesgue spaces
$L^{p}(\Omega,dm)$, their norms or pseudo norms. \ 

\section{The generalized R\'{e}nyi entropies (GRE).}

\subsection{Definitions and notations.}

We follow the same approach that from ref.\cite{SonninoSteinbrGRE},
\cite{GySGSgrestability}. We will define the Generalized R\'{e}nyi entropies
by using the results on Banach spaces with the anisotropic norm, exposed in
ref.\cite{Besov}. In the following we will restrict our discussions to the set
of parameters that define the GRE, when a) The integrals that appears in the
definition can be interpreted like distance in a suitable function space and
b) The formula for entropy can be related to convexity or concavity properties
of some functional, in the subspace of non negative density functions.
\ Consequently we will define only two class of distance functionals and
entropies, in analogy to the functionals $S_{p_{y},p_{z}}^{(1)}[\rho]$ and
$S_{q_{y},q_{z}}^{(2)}[\rho]$ defined in ref.\cite{SonninoSteinbrGRE}.

Consider that the measure space $(\Omega,\mathcal{A},m)$ has the following
product structure. \ The phase space $\Omega$ is split in $3$ subspaces%
\begin{equation}
\Omega=\Omega_{1}\times\Omega_{2}\times\Omega_{3} \label{LL9}%
\end{equation}

That means that the argument $\mathbf{x}$ of probability density function can
be represented as $\mathbf{x=}\left\{  x_{1},x_{2},x_{3}\right\}  $, so
\begin{equation}
\rho(\mathbf{x})=\rho(x_{1},x_{2},x_{3}) \label{LL9.1}%
\end{equation}

with $x_{k}\in\Omega_{k}$. We mention also that in general the component
spaces $\Omega_{k}$ has the structure of $\mathbf{R}^{n}$ or more general
infinite dimensional measure space. \ Each of the spaces $\Omega_{k}$ has
their $\sigma-$algebra $\mathcal{A}_{k}$. The $\sigma-$algebra $\mathcal{A}$,
that contains subsets of $\Omega=\Omega_{1}\times\Omega_{2}\times
...\times\Omega_{N}$ is defined as a tensor product: it is the largest
$\sigma-$algebra on $\Omega$ such that \ all of the projections $\Omega
\overset{p_{k}}{\rightarrow}\Omega_{k}$ are measurable.

The measure $m$ is also Factorizable:%
\begin{equation}
dm(\mathbf{x)=}dm(x_{1},x_{2},x_{3}\mathbf{\mathbf{)}=}\prod\limits_{j=1}%
^{3}dm_{k}(x_{k}) \label{LL10}%
\end{equation}

where the measures $m_{k}$ are defined on the $\sigma-$algebras $\mathcal{A}%
_{k}$.

In other words, the measure space $(\Omega,\mathcal{A},m)$ is the \ tensor
product\
\begin{equation}
(\Omega,\mathcal{A},m)=\bigotimes\limits_{j=1}^{3}(\Omega_{j},\mathcal{A}%
_{j},m_{j}) \label{LL11}%
\end{equation}

The elementary probability $dP(\mathbf{x})$ is given by%
\begin{equation}
dP(\mathbf{x})=\rho(x_{1},x_{2},x_{3})dm(\mathbf{x}) \label{LL12}%
\end{equation}

where $dm(\mathbf{x})$ is given by Eq.(\ref{LL10}).

Consider a vector $\mathbf{p}=\{p_{1},p_{2},p_{3}\}$ of \ real numbers with
$p_{k}\geq1$. According to Ref.\cite{Besov}, in close analogy to
Ref.\cite{SonninoSteinbrGRE} (where the particular case $N=2$ was studied) and
\cite{GySGSgrestability}, we define recursively the norm (depending on the
measure $m$) $\left\Vert \rho\right\Vert _{\mathbf{p},m}$ as follows%
\begin{align}
\rho_{2}(x_{1,},x_{2})  &  :=\left[  \int\limits_{\Omega_{N}}\left[
\rho_{\text{ }}(x_{1},x_{2},x_{3})\right]  ^{p_{3}}dm_{3}(x_{3})\right]
^{1/p_{3}}\label{LL13}\\
\rho_{1\text{ }}(x_{1})  &  :=\left[  \int\limits_{\Omega_{2}}\left[
\rho_{2\text{ }}(x_{1},x_{2})\right]  ^{p_{2}}dm_{2}(x_{2})\right]  ^{1/p_{2}%
}\ \label{LL16}\\
\left\Vert \rho\right\Vert _{\mathbf{p},m}  &  :=\left[  \int\limits_{\Omega
_{1}}\left[  \rho_{1\text{ }}(x_{1})\right]  ^{p_{1}}dm_{1}(x_{1})\right]
^{1/p_{1}}\ \label{LL17}%
\end{align}

In analogy with Eqs.(\ref{LL3}, \ref{LL5} ) and Ref.\cite{SonninoSteinbrGRE}
we define the GRE, with respect to the measure $m$
\begin{equation}
S_{\mathbf{p}}^{(1)}[\rho,m]=\frac{p_{1}}{1-p_{3}}\log\left\Vert
\rho\right\Vert _{\mathbf{p},m};p_{i}>1 \label{LL18}%
\end{equation}

Observe that the anisotropic norm function $\rho\rightarrow\left\Vert
\rho\right\Vert _{\mathbf{p},m}$ \ is convex and satisfies the axioms of norm.
The corresponding normed vector space is complete, i. e. it is a Banach space.
See ref.\cite{Besov}. \ There is another range of parameters that generalize
the R\'{e}nyi entropy corresponding to Eqs.(\ref{LL4}, \ref{LL6}). Consider a
vector $\mathbf{q}=\{q_{1},q_{2},q_{3}\}$ of \ real numbers with $0<q_{k}%
\leq1$. In analogy to Eqs.(\ref{LL13}-\ref{LL17}) we define recursively
\cite{GySGSgrestability}%

\begin{align}
\rho_{2}^{\prime}(x_{1},x_{2})  &  :=\int\limits_{\Omega_{N}}\left[
\rho(x_{1},x_{2},x_{3})\right]  ^{q_{3}}dm_{3}(x_{3}).\label{LL20}\\
\rho_{1\text{ }}^{\prime}(\mathbf{x}_{1})  &  :=\int\limits_{\Omega_{2}%
}\left[  \rho_{2\text{ }}^{\prime}(x_{1},x_{2})\right]  ^{q_{2}}dm_{2}%
(x_{2})\label{LL24}\\
N[\rho]_{\mathbf{q},m}  &  :=\int\limits_{\Omega_{1}}\left[  \rho_{1\text{ }%
}^{\prime}(x_{1})\right]  ^{q_{1}}dm_{1}(x_{1}) \label{LL25}%
\end{align}

Observe that the mapping $\rho\rightarrow N[\rho]_{\mathbf{p},m}$ defines a
pseudonorm \ on the space of probability density functions. The map
$\rho\rightarrow N[\rho]_{\mathbf{p},m}$ defines a concave function, in the
subset of physically admissible PDF's, when $\rho(x_{1},x_{2},x_{3})\geq0$.
The GRE\ will be defined in analogy to Eqs.(\ref{LL4}, \ref{LL6}) \ and to the
case $N=2$ from ref. \cite{SonninoSteinbrGRE}, \cite{GySGSgrestability}
\begin{equation}
S_{\mathbf{q}}^{(2)}[\rho,m]=\frac{1}{1-q_{3}}\log N[\rho]_{\mathbf{q}%
,m};0<q_{i}<1 \label{LL26}%
\end{equation}

For simplification of the notations we will use the extrapolated for of the
Eq.(\ref{LL26}) also to the range of parameters $\{q_{1},q_{2},q_{3}\}$ that
allows to relate $S_{\mathbf{q}}^{(2)}[\rho,m]$ to $S_{\mathbf{p}}^{(1)}%
[\rho,m]$. We obtain%
\begin{align}
S_{\mathbf{q}}^{(2)}[\rho,m]  &  =S_{\mathbf{p}}^{(1)}[\rho,m]\label{LL27}\\
N[\rho]_{\mathbf{q},m}  &  =\left[  \left\Vert \rho\right\Vert _{\mathbf{p}%
,m}\right]  ^{p_{1}} \label{LL28}%
\end{align}
when $p_{i}$ and $q_{i}$ are related as follows%
\begin{align}
q_{3}  &  =p_{3}\label{LL29}\\
q_{2}  &  =\frac{p_{2}}{p_{3}}\label{LL30}\\
q_{1}  &  =\frac{p_{1}}{p_{2}} \label{LL32}%
\end{align}

\begin{remark}
\label{RemDomain_p_q} The algebraic equations associated to maximal entropy
problem are very complicated in the general case, nevertheless from the
convexity or concavity properties we have some informations. According to
Eqs.(\ref{LL27}-\ref{LL32}), we are in the domain when $\rho\rightarrow
\left\Vert \rho\right\Vert _{\mathbf{p},m}$ is a convex functional when
\begin{equation}
p_{k}=\prod\limits_{j=k}^{3}q_{j}\geq1;\ 1\leq k\leq3 \label{LL33}%
\end{equation}
In this case the problem of maximal entropy with linear restriction \ is
equivalent to minimization of a positive convex function and has unique
solution. In the domain $0<q_{k}<1$, where the map $\rho\rightarrow
N[\rho]_{\mathbf{q,}m}$ \ is a concave function, the maxent problem is
equivalent with the maximization of a concave function with linear
restriction. If the solution exists it is unique.
\end{remark}

\subsection{Particular cases.}

In the following we will not omit the measure, when no confusion arise:
$\left\Vert \rho\right\Vert _{\mathbf{p},m}:=\left\Vert \rho\right\Vert
_{\mathbf{p}}$; $N[\rho]_{\mathbf{q},m}:=N[\rho]_{\mathbf{q}}$;
\ $S_{\mathbf{p}}^{(a)}[\rho,m]:=S_{\mathbf{p}}^{(a)}[\rho]$.

In the particular case when $p_{1}=p_{2}=p_{3}>1$, or \ \ $q_{1}=q_{2}=1$,
$~\ 0\,<\,q_{3}<1$ the GRE is equal to the classical R\'{e}nyi entropy from
Eqs.(\ref{LL5}, \ref{LL6}):%
\begin{align}
S_{\mathbf{p}}^{(1)}[\rho]  &  =S_{R,p_{3}}[\rho]=\frac{p_{3}}{1-p_{3}}%
\log\left\Vert \rho\right\Vert _{p_{3}}\label{LL52}\\
\left\Vert \rho\right\Vert _{p_{3}}  &  =\left[  \int\limits_{\Omega
}dm(\mathbf{x})\rho(\mathbf{x})^{p_{3}}\right]  ^{1/p_{3}} \label{LL53}%
\end{align}

respectively
\begin{align}
S_{\mathbf{q}}^{(2)}[\rho]  &  =S_{R,q_{3}}[\rho]=\frac{1}{1-q_{3}}\log
N[\rho]_{q_{3}}\label{LL54}\\
N[\rho]_{q_{3}}  &  =\int\limits_{\Omega}dm(\mathbf{x})\rho(\mathbf{x}%
)^{q_{3}} \label{LL55}%
\end{align}

We used the notation from Eqs.(\ref{LL3}, \ref{LL4}, \ref{LL16}, \ref{LL26}).
In particular when $p_{N}\searrow1$ in Eqs. (\ref{LL52}, \ref{LL53}),
respectively when $q_{1}=q_{2}=q_{N-1}=1;~q_{N}\nearrow1$ \ in Eqs.(\ref{LL54}%
, \ref{LL55}), we obtain the classical Boltzmann-Shannon entropy%
\begin{equation}
\underset{p_{1}=p_{2}=p_{3}\searrow1}{\lim}S_{\mathbf{p}}^{(1)}[\rho
]=\underset{q_{1}=q_{2}=1;~q_{3}\nearrow1}{\lim}S_{\mathbf{q}}^{(2)}%
[\rho]==-\int\limits_{\Omega}dm(\mathbf{x})\rho(\mathbf{x})\log\rho
(\mathbf{x}) \label{LL56}%
\end{equation}

Remark that the path to the limiting classical case is essential%
\begin{equation}
\underset{~q_{3}\nearrow1}{\lim}~\underset{q_{1}=q_{2}\nearrow1}{\lim
}S_{\mathbf{q}}^{(2)}[\rho]=-\int\limits_{\Omega}dm(\mathbf{x})\rho
(\mathbf{x})\log\rho(\mathbf{x}) \label{LL56.01}%
\end{equation}

while
\[
\underset{q_{1}=q_{2}\nearrow1}{\lim}~\underset{~q_{3}\nearrow1}{\lim
}~S_{\mathbf{q}}^{(2)}[\rho]=\infty
\]

\subsection{Geometric properties of GRE}

Despite for the physical application the previous definitions of the R\'{e}nyi
entropy and GRE are more advantageous, it is important to remark that in the
our geometric approach the basic objects are the norms defined in
Eqs.(\ref{LL13}-\ref{LL17}) or pseudo norms defined in Eqs.(\ref{LL20}%
-\ref{LL25}). \ 

In the case when $p_{k}\geq1$, $\mathbf{p}=\{p_{1},...,p_{N}\}$, \ (see
\cite{Besov}) the norm $\left\Vert .\right\Vert _{\mathbf{p}}$ has the usual
properties: for $a\in\mathbf{R}$ we have $\left\Vert a\rho\right\Vert
_{\mathbf{p}}=\left\vert a\right\vert \left\Vert a\rho\right\Vert
_{\mathbf{p}}$, respectively%
\begin{equation}
\left\Vert \rho_{1}+\rho_{2}\right\Vert _{\mathbf{p}}\leq\left\Vert \rho
_{1}\right\Vert _{\mathbf{p}}+\left\Vert \rho_{2}\right\Vert _{\mathbf{p}}.
\label{LL56.1}%
\end{equation}
In particular it follows the convexity of the mapping $\rho\rightarrow
\left\Vert \rho\right\Vert _{\mathbf{p}}$ : for $0\leq\alpha\leq1$ we have \
\begin{equation}
\left\Vert \alpha\rho_{1}+\left(  1-\alpha\right)  \rho_{2}\right\Vert
_{\mathbf{p}}\leq\alpha\left\Vert \rho_{1}\right\Vert _{\mathbf{p}}+\left(
1-\alpha\right)  \left\Vert \rho_{2}\right\Vert _{\mathbf{p}} \label{LL57}%
\end{equation}

In the case $0<q_{k}\leq1$, $\mathbf{q}=\{q_{1},...,q_{N}\}$, the properties
of the pseudo norms $N[\rho]_{\mathbf{q},m}$ defined in Eqs.(\ref{LL20}%
-\ref{LL25}) also allows geometrical interpretations. We have%
\begin{equation}
N[\rho_{1}+\rho_{2}]_{\mathbf{q}}\leq N[\rho_{1}]_{\mathbf{q}}+N[\rho
_{2}]_{\mathbf{q}} \label{LL58}%
\end{equation}
e. This can be seen by using t the definition and the simple inequality
$|x+y|^{q}\leq|x|^{q}+|y|^{q}$, with $0<q\leq1$. Instead of convexity we have
the following concavity inequality \textbf{ "in the first octant" only} : when
$\rho_{1,2}\geq0$%
\begin{equation}
N[\alpha\rho_{1}+\left(  1-\alpha\right)  \rho_{2}]_{\mathbf{q}}\geq\alpha
N[\rho_{1}]_{\mathbf{q}}+(1-\alpha)N[\rho_{2}]_{\mathbf{q}} \label{LL59}%
\end{equation}

that can be proven easily by using the concavity of the function $f(x):=x^{q}$
with $0<q\leq1$.

By defining, the distance function between distribution functions $\rho_{1}$
and $\rho_{2}$ in the infinite dimensional space of PDF's by $d(\rho_{1}%
,\rho_{2}):=\left\Vert \rho_{1}-\rho_{2}\right\Vert _{\mathbf{p}}$ for
$p_{k}\geq1$ respectively $d(\rho_{1},\rho_{2}):=N[\rho_{1}-\rho
_{2}]_{\mathbf{q}}$ for $0\,\,<q_{k}\leq1$, we have the triangle inequality
\begin{equation}
d(\rho_{1},\rho_{3})\leq d(\rho_{1},\rho_{2})+d(\rho_{2},\rho_{3})
\label{LL59.1}%
\end{equation}

that allows geometrical interpretation of GRE in term of distance in the
functional space of admissible PDF's.

\section{Maximal Generalized Entropy distributions \ \ }

The main objective of the our work is to obtain a shortest derivation of the
RDF for the charged particle distribution, studied in the previous works
\cite{sg1}-\cite{SGSGrRDFderivation}. We mention that by using the classical
MaxEnt principle in \cite{SGSGrRDFderivation} we obtained new derivation of
the RDF studied in \cite{sg1}-\cite{sg3}.

We consider, in the framework of the notations from Eqs.(\ref{LL10},
\ref{LL11}) the following problem: In the convex set $\mathcal{K}$ of PDF,
defined by the linear constraints and non-negativity condition%
\begin{align}
\int_{\Omega}dm(\mathbf{x})\rho(\mathbf{x})f_{k}(\mathbf{x})  &  =c_{k};~0\leq
k\leq M\label{LL60}\\
\rho(\mathbf{x})  &  \geq0 \label{LL61}%
\end{align}

find the PDF with maximal entropy a) $S_{\mathbf{p}}^{(1)}[\rho]$ with
$\ p_{k}>1$, or b) $S_{\mathbf{q}}^{(2)}[\rho]$, with $0<q_{k}<1$. In the
constraints we included also the normalization: $f_{0}(\mathbf{x}):=1$ and
$c_{0}=1$. \ \ In the case a) according to Eq.(\ref{LL18}) the MaxEnt problem
is equivalent convex optimization problem: to find the point \ on
$\mathcal{K}$ that the closest, to the origin, \textbf{in the sense of the
distance defined by the norm} $\left\Vert \rho\right\Vert _{\mathbf{p},m}$.
\ Due to the convexity of the norm Eq.(\ref{LL57}) and convexity of the set
$\mathcal{K}$, the solution of the maxent problem is unique. The existence is
related to converge problems \ If it exists then the solution is unique.,
irrespective how complicated are the algebraic equations for Lagrange
multipliers. In the case when the measure is finite, namely when $\int
_{\Omega}dm(\mathbf{x})<\infty$ and the functions \ are $f_{k}(\mathbf{x})$
are all bounded, then it is possible to find a set of parameters $c_{k}$ such
that the MaxEnt problem has a solution

In the case b) from Eq.(\ref{LL26}) results that the MaxEnt problem is to find
the \ point \ t $\mathcal{\rho\in K}$ \ with maximal value of $N[\rho
]_{\mathbf{q}}$. We observe that the distance function $d(\rho,\mathbf{0}%
):=N[\rho]_{\mathbf{q}}$ unlike to the familiar Euclidian distance is concave
\ for $\rho\geq0$, not convex. From the convexity of $\mathcal{K}$ \ and the
concavity inequality Eq. (\ref{LL59}) results that also in case b) the
solution of the maxent problem exists. \ 

\subsection{Explicit MaxEnt distributions for $N=3$}

\ 

In the previous works \cite{sg1}-\cite{sg3} \ we obtained realistic
PDF\ starting from classical or generalized MaxEnt principle with scale
invariant restrictions. \ Apparently, by adding sufficient large number of
polynomial restrictions we can locally approximate any distribution function
in the case of classical MaxEnt principle, by using Shannon Entropy. Our
derivation of PDF \cite{sg1}-\cite{sg3}\ can be considered as a harmonic
analysis under the group of affine transformations. Considering only
restrictions with lowest order polynomial means that in the harmonic analysis
we restrict ourselves to the lowest dimensional representations.

We will study the stationary point aspect in the MaxEnt problem for the GRE
defined by $S_{\mathbf{q}}^{(2)}[\rho]$ for $0<q_{k}<1$ as well as for the
domain defined by Eqs.(\ref{LL33}) that corresponds to the MaxEnt problem for
$S_{\mathbf{p}}^{(1)}[\rho]$, for $p_{k}>1$ (See Remark \ref{RemDomain_p_q}).

\ We will concentrate on the constrained maximization problem in the case
$0<q_{k}<1,~1\leq k\leq3$. Recall that we use the notations: $\mathbf{x}%
=(x_{1},x_{2},x_{3})$, $\mathbf{q}=(q_{1},q_{2},q_{3})$, $dm(\mathbf{x}%
)=dm_{1}(x_{1})dm_{2}(x_{2})dm_{3}(x_{3})$, $\Omega=\Omega_{1}\times\Omega
_{2}\times\Omega_{3}$

\bigskip Denote by $\mathbf{\lambda}=(\lambda_{0},...,\lambda_{M})\ $\ the
Lagrange multipliers associated to restrictions Eq.(\ref{LL60}) and by
$\mu(\mathbf{x)}$ the multiplier associated to the restriction Eq.(\ref{LL61}%
). From Kuhn-Tucker theorem for maximization \cite{KuhnTucker}, we get
\begin{align}
\frac{\delta}{\delta\rho(\mathbf{x})}\left\{  N_{\mathbf{q}}[\rho
]+\!\!\!\!\!{\int\limits_{\Omega}}dm(\mathbf{x)}\rho\mathbf{(x)}\left[
\mu(\mathbf{x})-{\sum\limits_{k=0}^{M}}\lambda_{k}f_{k}(\mathbf{x})\right]
\right\}   &  =0\label{LL62}\\
\mu(\mathbf{x})  &  \geq0;~\mu(\mathbf{x})\rho\mathbf{(x)=0} \label{LL63}%
\end{align}

We introduce the following notations%
\begin{align}
g(\mathbf{\lambda,x})  &  :=\frac{1}{\ q_{1}q_{2}q_{3}}{\sum\limits_{k=0}^{N}%
}\lambda_{k}f_{k}(\mathbf{x})\label{LL64}\\
k_{3}(\mathbf{\lambda,x})  &  :=\left[  g(\mathbf{\lambda,x})\right]
_{+}^{1/\left(  q_{3}-1\right)  } \label{LL65}%
\end{align}

We will use here and in the Appendix \ref{markerAppendixDerivationExtremalPDF}
the following notations for exponents. \
\begin{align}
c_{2}  &  =\frac{1-q_{2}}{q_{2}q_{3}-1}\label{LL65.1}\\
d_{2}  &  =\ \frac{\ q_{2}(q_{3}-1)}{q_{2}q_{3}-1}\label{LL65.2}\\
c_{1}  &  =\frac{1-q_{1}}{q_{1}q_{2}q_{3}-1} \label{LL65.3}%
\end{align}
We also denote%
\begin{align}
k_{2}(\mathbf{\lambda},x_{1},x_{2})  &  :=\int\limits_{\Omega_{3}}dm_{3}%
(x_{3}^{\prime})k_{3}(\mathbf{\lambda},x_{1},x_{2},x_{3}^{\prime})^{q_{3}%
}\label{LL66}\\
k_{1}(\mathbf{\lambda},x_{1})  &  :=\int\limits_{\Omega_{2}}dm_{2}%
(x_{2}^{\prime})\left[  k_{2}(\mathbf{\lambda},x_{1},x_{2}^{\prime})\ \right]
^{d_{2}}; \label{LL67}%
\end{align}

The distribution function that satisfy the stationarity condition
\ Eq.(\ref{LL62}) is (for details see Appendix
\ref{markerAppendixDerivationExtremalPDF})
\begin{equation}
\rho(\mathbf{\lambda},\mathbf{x})=k_{3}(\mathbf{\lambda},x_{1},x_{2}%
,x_{3})k_{2}(\mathbf{\lambda},x_{1},x_{2})^{c_{2}}\left[  k_{1}%
(\mathbf{\lambda},x_{1})\right]  ^{c_{1}} \label{LL67.1}%
\end{equation}

\section{ \ \ Examples\ }

\subsection{Symmetric distributions}

Consider the following examples: $\Omega=\Omega_{1}\times\Omega_{2}%
\times\Omega_{3}$ \ with $\Omega_{1}=\Omega_{2}=\Omega_{3}=\mathbb{R}$,
$dm_{k}(x_{k})=dx_{k}$ for $1\leq k\leq3$ and for the restrictions from
Eq.(\ref{LL60}) we have
\begin{align}
f_{0}(x_{1},x_{2},x_{3})  &  :=1\label{LL69}\\
f_{k}(x_{1},x_{2},x_{3})  &  :=x_{k}^{2};~1\leq k\leq3 \label{LL70}%
\end{align}

\bigskip{}

\bigskip%
\[
f_{k}(x_{1},x_{2},x_{3}):=x_{k}^{2};~1\leq k\leq3
\]

By using Eqs.(\ref{LL64}-\ref{LL67.1}) we obtain the MaxEnt distribution in
the case $0<q_{k}<1$ as follows ( $C_{1,2,3}$ are constants)%
\begin{align}
k_{2}(\mathbf{\lambda},x_{1},x_{2})  &  =\frac{C_{2}}{\left(  1+a_{1}^{2}%
x_{1}^{2}+a_{2}^{2}x_{2}^{2}\right)  ^{r_{2}}\ }~\ \label{LL70.1}\\
r_{2}  &  =\frac{q_{3}}{1-q_{3}}-\frac{1}{2}>0~~~~\label{LL70.2}\\
k_{1}(\mathbf{\lambda},x_{1})  &  =\frac{C_{1}}{\left(  1+a_{1}^{2}x_{1}%
^{2}\ \right)  ^{r_{1}}}~~~~\label{LL70.3}\\
r_{1}  &  =n_{3}r_{2}-\frac{1}{2}>0~~~~ \label{LL70.4}%
\end{align}

From Eqs.(\ref{LL67.1}, \ref{LL70.1}, \ref{LL70.3}) results
\begin{equation}
\rho(\mathbf{\lambda},\mathbf{x})=a_{0}\left(  1+a_{1}^{2}x_{1}^{2}\right)
^{b_{1}}(1+a_{1}^{2}x_{1}^{2}+a_{2}^{2}x_{2}^{2})^{b_{2}}\left(  1+a_{1}%
^{2}x_{1}^{2}+a_{2}^{2}x_{2}^{2}+a_{3}^{2}x_{3}^{2}\right)  ^{-b_{3}}
\label{LL71}%
\end{equation}
where $a_{k}$ are free parameters, and the exponents $b_{k}$ are given by
\begin{align}
b_{1}  &  =-r_{1}n_{4}=r_{1}\frac{1-q_{1}}{1-q_{1}q_{2}q_{3}}%
\ \ \ \ \ \ \ \label{LL72}\\
b_{2}  &  =-r_{2}n_{1}=\left(  \ \frac{q_{3}}{1-q_{3}\ }-\frac{1}{2}\right)
\frac{1-q_{2}}{1-q_{2}q_{3}}\ \ \ \ \label{LL73}\\
b_{3}  &  =\frac{1}{1-q_{3}\ }\ \ \ \ \ \ \label{LL74}%
\end{align}

The restrictions for $b_{1,2,3}$ resulting from the finiteness of
$\left\langle x_{k}^{2}\right\rangle ~\ $\ are
\begin{align}
b_{3}-\frac{3}{2}  &  >0~~~\label{LL75}\\
b_{3}-b_{2}-2  &  >0~~~~\label{LL76}\\
b_{3}-b_{1}-b_{2}-\frac{5}{2}~  &  >0~ \label{LL77}%
\end{align}

By simple but tedious algebra ( we used the Reduce command from MATHEMATICA
5.1 ) \ the domain given by Eqs. (\ref{LL70.2}, \ref{LL70.4}, \ref{LL72}%
-\ref{LL74}), with restriction $0<q_{k}<1$, the resulting domain in the
variables $q_{1},q_{2},q_{3}$ is given by
\begin{align}
1/3  &  <q1<1\label{LL78}\\
\frac{1+q_{1}}{4q_{1}}  &  <q_{2}<1\label{LL79}\\
\frac{1+q_{1}+q_{1}q_{2}}{5q_{1}q_{2}}  &  <q_{3}\ <1 \label{LL80}%
\end{align}
It is easy to verify that the set defined by Eqs(\ref{LL78}-\ref{LL80}) is not
empty: it is an open set and contains at least some open neighborhood of the
point defined by $(q_{1},q_{2},q_{3})=(101/152,267/271,262/347)$. It can be
proven that in the limit $q_{k}\nearrow1$ and $a_{k}\searrow0$ \ with suitable
scaling we obtain the centered Gaussian distribution, in a similar manner to
the following example.

\subsection{ The RDF for charged particle distribution, derived from
Generalized Maximal Entropy principle}

\subsubsection{The RDF obtained from classical MaxEnt principle \cite{sg1}-\cite{SGSGrRDFderivation}}

The reference state studied in \cite{sg1}-\cite{SGSGrRDFderivation} has the
form
\begin{align}\label{11-ddf}
d\widehat{\mathcal{F}}^{R}=&\ \mathcal{N}_{0}\left(  \frac{w}{\Theta}\right)
^{\gamma-1}\exp\left[  -w/\Theta\right]  \exp\left[  -c_{1}\left(
w/\Theta\right)  \left(  P_{\phi}-P_{\phi0}\right)  ^{2}\right] \nonumber\\
&\exp\left[
-c_{2}\left(  w/\Theta\right)  \left(  \lambda-\lambda_{0}\right)
^{2}\right]  \left\vert \mathcal{J}\right\vert d\widehat{\Gamma}
\end{align}
where $P_{\phi},\lambda$ and $w$ are the invariants appearing in the axial
symmetric magnetic field variables \cite{sg1}-\cite{SGSGrRDFderivation}
$\ \Theta$ is the scale parameter for energy, $\gamma$ the shape parameter of
the gamma distribution that appears in Eq.(\ref{11-ddf}). Supposing that
\begin{align*}
c_{1}\left(  w/\Theta\right)   &  \simeq c_{1}^{\left(  0\right)  }%
\equiv\left(  \frac{1}{\Delta P_{\phi}}\right)  ^{2}=const.\\
c_{2}\left(  w/\Theta\right)   &  \simeq c_{2}^{\left(  0\right)  }%
+c_{2}^{\left(  1\right)  }\frac{w}{\Theta}\equiv\frac{1}{\Delta\lambda_{0}%
}\left(  \frac{\Delta\lambda_{0}}{\Delta\lambda_{1}}+\frac{w}{\Theta}\right)
\geq0
\end{align*}
the density distribution function $\widehat{\mathcal{F}}^{R}$ reads as%
\begin{align}\label{14-ddf}
\widehat{\mathcal{F}}^{R}=&\ \mathcal{N}_{0}\left(  \frac{w}{\Theta}\right)
^{\gamma-1}\exp\left[  -w/\Theta\right]  \exp\left[  -\left(  \frac{P_{\phi
}-P_{\phi0}}{\Delta P_{\phi}}\right)  ^{2}\right]\nonumber\\
&\exp\left[  -\left(
\frac{\Delta\lambda_{0}}{\Delta\lambda_{1}}+\frac{w}{\Theta}\right)
\frac{\left(  \lambda-\lambda_{0}\right)  ^{2}}{(\Delta\lambda_{0})^{2}
}\right]  \left\vert \mathcal{J}\right\vert 
\end{align}
where $\Delta P_{\phi}$, $\Delta\lambda_{0}$ and $\Delta\lambda_{1}$ are
constants and $\mathcal{N}_{0}$ ensures normalization to unity.

\subsubsection{The RDF obtained from maximal generalized entropy principle}

We will study the case when $0<q_{k}<1$, that generate extremal distributions
with algebraic decay at infinity. Consider the following restrictions
\begin{align*}
f_{0}(x_{1},x_{2},x_{3})  &  \equiv1\\
f_{1}(x_{1},x_{2},x_{3})  &  \equiv\left\vert x_{1}\right\vert ^{\alpha_{1}}\\
f_{0}(x_{1},x_{2},x_{3})  &  \equiv\left\vert x_{1}\right\vert ^{\delta
}\left\vert x_{2}\right\vert ^{\alpha_{2}}\\
f_{0}(x_{1},x_{2},x_{3})  &  \equiv\left\vert x_{3}\right\vert ^{\alpha_{3}}%
\end{align*}
\noindent where $x_{1}\in\mathbb{R}_{+}$ and $x_{3},x_{2}\in\mathbb{R}$. We consider the
case of distributions defined in the whole phase space, so we select the
Lagrange multipliers strictly positive. \ According to Eqs.(\ref{LL64}%
-\ref{LL67.1}) as well as Eq.(\ref{int1}) results, up to irrelevant constant
factor \
\begin{align}\label{LL81}
&k_{3}(\mathbf{\lambda},\mathbf{x)} \mathbf{=}\left[  \lambda_{0}
+\lambda_{1}\left\vert x_{1}\right\vert ^{\alpha_{1}}+\lambda_{2}\left\vert
x_{1}\right\vert ^{\delta}\left\vert x_{2}\right\vert ^{\alpha_{2}}
+\lambda_{3}\left\vert x_{3}\right\vert ^{\alpha_{3}}\right]  _{+}
^{1/(q_{3}-1)}\\
&k_{2}(\mathbf{\lambda},x_{1},x_{2}\mathbf{)} \mathbf{=}\left[  \lambda
_{0}+\lambda_{1}\left\vert x_{1}\right\vert ^{\alpha_{1}}+\lambda
_{2}\left\vert x_{1}\right\vert ^{\delta}\left\vert x_{2}\right\vert
^{\alpha_{2}}\right]  ^{-m_{2}}\nonumber\\
&k_{1}(\mathbf{\lambda},x_{1}\mathbf{)} \mathbf{=}x_{1}^{-\delta/\alpha
_{2}}\left[  \lambda_{0}+\lambda_{1}\left\vert x_{1}\right\vert ^{\alpha_{1}
}\right]  ^{-m_{1}}\nonumber\\
&\rho(\mathbf{\lambda},\mathbf{x}) =x_{1}^{-r_{1}}\left[  \lambda
_{0}+\lambda_{1}\left\vert x_{1}\right\vert ^{\alpha_{1}}\right]  ^{-r_{2}
}\!\left[\lambda_{0}+\lambda_{1}\left\vert x_{1}\right\vert ^{\alpha_{1}
}+\lambda_{2}\!\left\vert x_{1}\right\vert ^{\delta}\left\vert x_{2}\right\vert
^{\alpha_{2}}\right]  ^{-r_{3}}\!\! k_{3}(\mathbf{\lambda},\mathbf{x)} 
\nonumber \end{align}
\noindent where we used the notations
\begin{align}
m_{2}  &  =\frac{q_{3}}{1-q_{3}}-\frac{1}{\alpha_{3}}>0\label{LL82}\\
m_{1}  &  =m_{2}d_{2}-\frac{1}{\alpha_{2}}>0\label{LL83}\\
r_{1}  &  =\frac{\delta}{\alpha_{2}}c_{1};~r_{2}=m_{1}c_{1}\label{LL84}\\
r_{3}  &  =m_{2}c_{2} \label{LL85}%
\end{align}
\noindent and Eqs.(\ref{LL65.1}-\ref{LL65.3}). We consider now the choice of the free parameters $\alpha_{i},\delta$,
$q_{i}$ in Eq.(\ref{LL67.1}) such that we recover a DDF from the family
Eqs.(\ref{11-ddf}, \ref{14-ddf}). Then according to Eq.(\ref{LL56.01}) we set
\begin{equation}
q_{k}=1-\varepsilon_{k};~\varepsilon_{k}\searrow0 \label{LL86}%
\end{equation}
\noindent In order to recover Eqs.(\ref{11-ddf}, \ref{14-ddf}) we set
\[
\alpha_{1}=1;~\alpha_{2}=\alpha_{3}=2
\]
\noindent without loss of generality we set $\lambda_{0}=1$. Up to linear terms in
$\varepsilon_{k}$ we obtain%
\begin{align*}
c_{2}  &  =-\frac{\varepsilon_{2}}{\varepsilon_{2}+\varepsilon_{3}%
}+\mathcal{O}(\varepsilon_{k})\\
d_{2}  &  =\frac{\varepsilon_{3}}{\varepsilon_{2}+\varepsilon_{3}}%
+\mathcal{O}(\varepsilon_{k})\\
c_{1}  &  =-\frac{\varepsilon_{1}}{\varepsilon_{1}+\varepsilon_{2}%
+\varepsilon_{3}}+\mathcal{O}(\varepsilon_{k})\\
m_{1}  &  =\frac{1}{\varepsilon_{2}+\varepsilon_{3}}+\frac{1}{2}%
\frac{\varepsilon_{3}}{\varepsilon_{2}+\varepsilon_{3}}+\mathcal{O}%
(\varepsilon_{k})>0\\
m_{2}  &  =\frac{q_{3}}{1-q_{3}}-\frac{1}{\alpha_{3}}=\frac{1}{\varepsilon
_{3}}-\frac{3}{2}%
\end{align*}
\noindent By using the notations
\begin{align*}
s_{2}  &  =\frac{\varepsilon_{2}\varepsilon_{3}}{(\varepsilon_{1}%
+\varepsilon_{2}+\varepsilon_{3})(\varepsilon_{2}+\varepsilon_{3})}\\
s_{3}  &  =\frac{\varepsilon_{3}}{\varepsilon_{2}+\varepsilon_{3}}%
\end{align*}
we have the following asymptotic form for the exponents in Eq.(\ref{LL81})
\begin{align*}
\ r_{3}  &  =-\frac{s_{3}}{\varepsilon_{3}}+\mathcal{O}(\varepsilon_{k})\\
r_{2}  &  =-\frac{s_{2}}{\varepsilon_{3}}+\mathcal{O}(\varepsilon_{k})\\
r_{1}  &  =-\frac{\delta\varepsilon_{1}}{\alpha_{2}(\varepsilon_{1}%
+\varepsilon_{2}+\varepsilon_{3})}+\mathcal{O}(\varepsilon_{k})
\end{align*}
\noindent Now we consider the limit $\varepsilon_{k}\rightarrow0$, with $\lambda_{0}=1$
\ and $s_{2}$, $s_{3}$, $r_{1}$ fixed and $\lambda_{k}\rightarrow0$ such that
\[
\lambda_{k}=\varepsilon_{3}\nu_{k};1\leq k\leq3
\]
and $\ \nu_{k};1\leq k\leq3$ \ are fixed positive constants. We obtain the
following \ result in the limit $\varepsilon_{3}\rightarrow0$%
\begin{equation}
\rho(\mathbf{\lambda},\mathbf{x})=x_{1}^{-r_{1}}\exp\left[  -\mu_{1}x_{1}%
-\mu_{2}x_{1}^{\delta}x_{2}^{2}-\mu_{3}x_{3}^{2}\right]  \label{LL87}%
\end{equation}
where the following notation was used%
\begin{align*}
\mu_{1}  &  =\nu_{1}\frac{\varepsilon_{2}}{(\varepsilon_{1}+\varepsilon
_{2}+\varepsilon_{3})}>0\\
\mu_{2}  &  =\nu_{2}\frac{\varepsilon_{2}}{(\varepsilon_{2}+\varepsilon_{3}%
)}>0\\
\mu_{3}  &  =\nu_{3}>0
\end{align*}
\noindent In order to approach the class of RDF from Eqs.(\ref{11-ddf}-\ref{14-ddf}), we
put in Eq.(\ref{LL87})
\begin{align*}
x_{1}  &  =w\\
x_{2}  &  =\lambda-\lambda_{0}\\
x_{3}  &  =P_{\Phi}-P_{\Phi0}\\
r_{1}  &  =1-\gamma<1\\
\frac{1}{\Theta}  &  =\mu_{1}\\
\frac{1}{(\Delta P_{\Phi})^{2}}  &  =\mu_{3}%
\end{align*}
\noindent and we obtain the limiting RDF of the form \cite{ditroia}
\begin{equation}\label{LL88}
\!\widehat{\mathcal{F}}^{R}=\mathcal{N}_{0}\left(\frac{w}{\Theta}\right)
^{\gamma-1}\!\!\!\!\exp\left[  -w/\Theta\right]  \exp\left[  -\left(  \frac{P_{\phi
}-P_{\phi0}}{\Delta P_{\phi}}\right)  ^{2}\right] \!\exp\left[  -w^{\delta
}\frac{\left(  \lambda-\lambda_{0}\right)  ^{2}}{(\Delta\lambda_{0})^{2}
}\right]  
\end{equation}
\noindent which excepting to a constant factor in $\frac{\left(  \lambda-\lambda
_{0}\right)  ^{2}}{(\Delta\lambda_{0})^{2}}$ term reproduces well the
qualitative behavior of the RDF from Eqs.(\ref{11-ddf}-\ref{14-ddf}).

\noindent However, it is correct to mention that the RDFs (\ref{11-ddf}, \ref{14-ddf}) and Eq.~(\ref{LL88}) should be understood as pure {\it pedagogical} distribution functions and not as realistic RDFs. Indeed, it is easily to check that there exist several ranges of parameters where these RDFs violate the energy balance equations. The interest of these RDFs resides solely on the fact that they can be used as RDF-fitting functions and that, as demonstrated in this work and in \cite{sg1}-\cite{SGSGrRDFderivation}, these fittings are derivable from the MaxEnt principle.

\section{Conclusions}

The nonlinear equations associated to the maximal Generalized R\'{e}nyi
Entropy problem was solved in the case of GRE with 3 variables. By simple
linear scale invariant restrictions a family of reference particle
distribution function for charged particles in tokamak was obtained.

\section{Appendix \label{MarkerAppendix}}

\subsection{Integrals \label{markerSugsectIntegrals}}

The following formula will be used%
\begin{equation}
\int\limits_{0}^{\infty}\frac{dx}{\left(  a+bx^{\alpha}\right)  ^{r}%
}=b^{-1/\alpha}a^{(1/\alpha-r)}\frac{1}{\alpha}B\left(  1/\alpha
,r-1/\alpha\right)  \label{int1}%
\end{equation}
\noindent where $a>0$, $\alpha>0$, $b>0$, $r>1/\alpha$.

\begin{proof}
Eq.(\ref{int1}) is reduced to the standard integral representation of the
Euler Beta function by the substitution%
\[
x=\left(  \frac{a}{b}\frac{u}{1-u}\right)  ^{1/\alpha}%
\]

\end{proof}

\subsection{Derivation of the extremal PDFs
\label{markerAppendixDerivationExtremalPDF}}

From Eqs.(\ref{LL62}, \ref{LL64}) \ we obtain the following nonlinear integral
equation for PDF $\rho(x_{1},x_{2},x_{3})$%
\begin{equation}
B(x_{1})^{q_{1}-1}A(x_{1},x_{2})^{q_{2}-1}\rho(x_{1},x_{2},x_{3})^{q_{3}%
-1}=g(\mathbf{\lambda},\mathbf{x}) \label{ap1}%
\end{equation}
\noindent where we denoted
\begin{align}
A(x_{1},x_{2})  &  =\int\limits_{\Omega_{3}}dm_{3}(x_{3}^{\prime})\rho
(x_{1},x_{2},x_{3}^{\prime})^{q_{3}}\label{ap2}\\
B(x_{1})  &  =\int\limits_{\Omega_{2}}dm_{2}(x_{2}^{\prime})\left[
A(x_{1},x_{2}^{\prime})\ \right]  ^{q_{2}} \label{ap3}%
\end{align}
\noindent In the following we will use the notations \ from Eqs.(\ref{LL64}-\ref{LL67}).
The solution of system of Eqs (\ref{ap1}-\ref{ap3}) is of the form
\begin{equation}
\rho(x_{1},x_{2},x_{3})=\rho_{2}(x_{1},x_{2})g(\mathbf{\lambda},\mathbf{x}%
)^{1/\left(  q_{3}-1\right)  }=\rho_{2}(x_{1},x_{2})k_{3}(\mathbf{\lambda
},\mathbf{x})\ \label{ap4}%
\end{equation}
\noindent where $\rho_{2}(x_{1},x_{2})$ remains to be determined. \ By inserting
Eq.(\ref{ap4}) in Eqs.(\ref{ap1}, \ref{ap2}) we obtain successively (see
notations \ from Eqs.(\ref{LL66}, \ref{LL67}))%
\begin{align}
B(x_{1})^{q_{1}-1}A(x_{1},x_{2})^{q_{2}-1}\rho_{2}(x_{1},x_{2})^{q_{3}-1}  &
=1\label{ap5}\\
A(x_{1},x_{2})  &  =\rho_{2}(x_{1},x_{2})^{q_{3}}k_{2}(\mathbf{\lambda}%
,x_{1},x_{2}) \label{ap6}%
\end{align}
\noindent We will use the notations for exponents \
\begin{align}
c_{1}  &  =\frac{1-q_{1}}{q_{1}q_{2}q_{3}-1}\label{ap7}\\
d_{2}  &  =\ \frac{\ q_{2}(q_{3}-1)}{q_{2}q_{3}-1}\label{ap8}\\
c_{2}  &  =\frac{1-q_{2}}{q_{2}q_{3}-1}\label{ap9}\\
n_{2}  &  =\frac{q_{3}-1}{q_{2}q_{3}-1}%
\end{align}
\noindent From Eqs.(\ref{ap5}, \ref{ap6}) results%
\begin{align}
\rho_{2}(x_{1},x_{2})  &  =\rho_{1}(x_{1})k_{2}(\mathbf{\lambda},x_{1}%
,x_{2})^{c_{2}}\label{ap10}\\
A(x_{1},x_{2})  &  =\rho_{1}(x_{1})^{q_{3}}k_{2}(\mathbf{\lambda},x_{1}%
,x_{2})^{n_{2}} \label{ap11}%
\end{align}
\noindent where $\rho_{1}(x_{1})$ remains to be found. Now we insert Eqs.(\ref{ap11},
\ref{ap10}) in Eqs.( \ref{ap3}, \ref{ap5}), we obtain%
\begin{align}
B(x_{1})  &  =\rho_{1}(x_{1})^{q_{2}q_{3}}k_{1}(\mathbf{\lambda}%
,x_{1})\label{ap12}\\
B(x_{1})^{q_{1}-1}\rho_{1}(x_{1})^{q_{2}q_{3}-1}  &  =1 \label{ap13}%
\end{align}
\noindent and from here we get
\begin{equation}
\rho_{1}(x_{1})=\left[  k_{1}(\mathbf{\lambda},x_{1})\right]  ^{c_{1}}
\label{ap14}%
\end{equation}
\noindent Collecting the previous results we obtain%
\begin{equation}
\rho(\mathbf{\lambda},\mathbf{x})=k_{3}(\mathbf{\lambda},x_{1},x_{2}%
,x_{3})k_{2}(\mathbf{\lambda},x_{1},x_{2})^{c_{2}}\left[  k_{1}%
(\mathbf{\lambda},x_{1})\right]  ^{c_{1}} \label{ap16}
\end{equation}

\end{document}